\documentclass[epsfig, proceedings]{rmaa}
\usepackage{graphicx}

\renewcommand{\P}[1]{%
\ifnum#1=1\hbox{OW~168--326E}\fi
\ifnum#1=2\hbox{OW~167--317}\fi
\ifnum#1=3\hbox{OW~163--317}\fi
\ifnum#1=5\hbox{OW~158--323}\fi
\ifnum#1=0\hbox{OW~171--334}\fi}

\title{Three Dimensional Radiative Transfer}
\author{Tom Abel\affil{CfA, Harvard}}

\fulladdresses{
\item Tom Abel: CfA, 60 Garden
Street, Cambridge, MA 02138, US ({\tt tabel@cfa.harvard.edu}) }

\shortauthor{Tom Abel}
\shorttitle{3D Radiative Transfer}


\abstract{Radiative Transfer (RT) effects play a crucial role in the
  thermal history of the intergalactic medium. Here I discuss recent
  advances in the development of numerical methods that introduce RT
  to cosmological hydrodynamics. These methods can also readily be
  applied to time dependent problems on interstellar and galactic
  scales.  }

\nonstopmode
\begin{document}

\maketitle

The physics of photoionized gases is of crucial importance and many
astrophysical applications. As the hydrodynamic modeling of
astronomical objects is advancing at a rapid pace we are also forced
to consider the complex problem of three dimensional radiative transfer of
ionizing radiation. I will report here on recent advances we have made
in developing appropriate numerical techniques to achieve this
goal. These methods are applicable for a wide range of problems. One
particular issue that arises in the formation of the first generation
of stars shall serve as motivation.

\section{One Motivation: Structure Formation and the beginning of the
bright ages}
\label{sec:intro}
 
The atomic nuclei in the primordial gas (mostly hydrogen and helium)
first (re)combined with electrons at a redshift $z\sim 1000$. From the
study of absorption spectra of high redshift quasars we know that this
then neutral gas must have been ionized prior to $z=5$. Most likely
this reionization process was caused via photoionization by UV photons
produced in proto galactic objects either by massive stars or by the
accretion onto compact objects. The formation of these first objects
in the universe and their potential impact on subsequent structure
formation is a highly topical issue in physical cosmology to date. In
our standard models of structure formation cosmological objects form
via hierarchical build up from smaller pieces. The dynamics is
controlled by gravity of a dominant cold dark matter (CDM)
component. Baryons will fall into virializing CDM halos in which they
may cool and possibly fragment to form stars.  The lower limit on the
masses of luminous objects that may be formed is determined by
(1)
 the pressure of the primordial gas, which determines whether it
can settle in the dark matter halo
(2) the ability of the baryons to cool to collapse to stellar densities.

Both these issues depend sensitively on the presence of UV photons
(see Abel and Haehnelt 1999, Haiman, Abel and Rees 1999, and
references therein). Since the intergalactic medium (IGM) is initially
optically thick to $h\nu>13.6$\,eV photons ionization fronts will be
formed around the first sources.  Because we believe that the
radiation is produced in structures condensed from the IGM by
gravitational instability, {\sl the first UV photons will see a clumpy
inhomogeneous IGM}. As a consequence the time--varying ionized regions
will have complex morphologies. 

The above physical processes have prompted us to develop methods for
the treatment of RT in three dimensional cosmological hydrodynamics.
In the following we describe the conditions under which the
cosmological RT becomes equivalent to classical RT. Then we go on to
discuss some possible methods to solve the latter. In this
contribution I will only give a brief overview of my and my
collaborators work. However, note that also Razoumov and Scott (1999)
offer a different approach. Gnedin (1999) chose to use dramatic
simplifications constructed such as to mimic the expected effects of
radiative transfer in order to study the reionization of intergalactic
hydrogen.

\section{Cosmological Radiative Transfer}
\label{sec:CRT}

The equation of cosmological radiative transfer in comoving coordinates
(cosmological, not fluid) is:

\begin{equation}
\frac{1}{c} \frac{\partial I_{\nu}}{\partial t} + \frac{\hat{n} \cdot
\nabla I_{\nu}}{\bar{a}} - \frac{H(t)}{c} (\nu \frac{\partial I_{\nu}}
{\partial \nu} - 3 I_{\nu}) = \eta_{\nu} - \chi_{\nu} I_{\nu}
\end{equation}
where $I_{\nu} \equiv I(t, \vec{x}, \vec{\Omega}, \nu)$ is the
monochromatic specific intensity of the radiation field, $\hat{n}$ is
a unit vector along the direction of propagation of the ray; $H(t)
\equiv \dot{a}/a$ is the (time-dependent) Hubble constant, and
$\bar{a} \equiv \frac{1+z_{em}}{1+z}$ is the ratio of cosmic scale
factors between photon emission at frequency $\nu$ and the present
time t.  The remaining variables have their traditional meanings (e.g,
Mihalas 1978.)  Equation (1) will be recognized as the standard
equation of radiative transfer with two modifications: the denominator
$\bar{a}$ in the second term, which accounts for the changes in path
length along the ray due to cosmic expansion, and the third term,
which accounts for cosmological redshift and dilution.

One could, in principle, attempt to solve equation (1) directly for
the intensity at every point given the emissivity $\eta$ and
absorption coefficient $\chi$. However, the high dimensionality of the
problem (three positions, two angles, one frequency and time $=$ 7D!)
not to mention the high spatial and angular resolution needed in
cosmological simulations would make this approach impractical for
dynamic computations. Therefore we proceed through a sequence of
well-motivated approximations which reduce the complexity to a
tractable level.

\subsection{Local quasi--static Approximation}

We begin by eliminating the cosmological terms and factors. That we
can do this can be understood on simple physical grounds. Before the
universe is reionized, it is opaque to H and He Lyman continuum
photons. Consequently, ionizing sources are local to scales of
interest, and not at cosmological distances. In particular this means
that the term multiplicative term $\frac{H(t)}{c}$ which is simply the
reciprocal of the Hubble horizon at the time $t$ will ensure the
cosmological term to be small as long as the opacity is much smaller
than the horizon scale. If this is not the case and we have a
simulation box size much smaller than the mean free path than we are
at the limit where the cosmological terms will modify the boundary
values but still not be important as the photons transverse the box.
Therefore, setting $\bar{a} \equiv 1$, equation (1) reduces to its
standard, non-cosmological form:

\begin{equation}\label{equ:clt}
\frac{1}{c} \frac{\partial I_{\nu}}{\partial t} + \hat{n} \cdot
\nabla I_{\nu} = \eta_{\nu} - \chi_{\nu} I_{\nu}
\end{equation}
\noindent
where now $\nu$ is the instantaneous, comoving frequency.

Thinking of the special case of a point source that switches on
instantaneously one realizes that initially the ionization front will
always propagate at the speed of light. Eventually it slows down as
the 'incoming flux' of neutrals grows with the increasing ionization
surface. This will ensure that eventually the light crossing time
(1/(c times I-front distance)) will become much shorter than the
timescales of change of the emissivities and absorption coefficient on
the right hand side of equation (\ref{equ:clt}). At this point an
explicit integration can employ large time steps making the term
$\frac{1}{c} \frac{\partial I_{\nu}}{\partial t}$ negligible. From
this point of the evolution on it will suffice to solve the static
classic equation of radiative transfer ,
\begin{equation}
\hat{n} \cdot \nabla I_{\nu} = \eta_{\nu} - \chi_{\nu} I_{\nu}. 
\end{equation}

\section{Methods}
\subsection{Brute Force: Ray Tracing}
Abel, Norman and Madau (1999) give a method that integrates this
quasi--static approximation along rays casted from point--sources.
That method has the particular advantage that it will ensure photon
conservation independent of resolution by exploiting the known analytic
solution of radiative transfer for a homogeneous slab.
\begin{figure}
  \begin{center} \leavevmode
    \includegraphics[width=7cm]{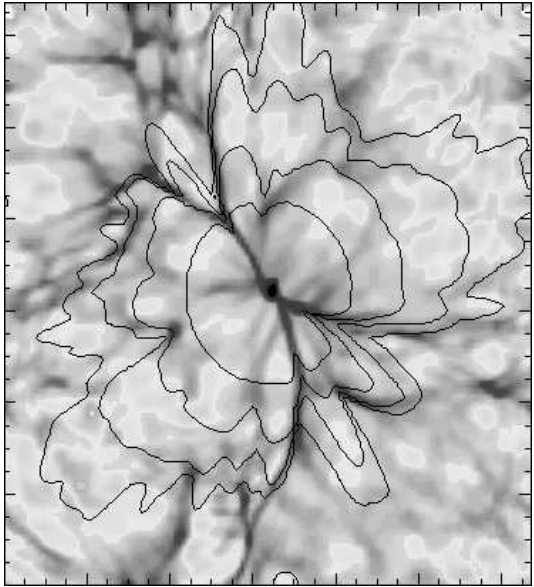}
    \caption{Propagation of an R-type I--front in a $128^3$
    cosmological density field produced by a mini-quasar with $\dot
    {N}=5\times 10^{53}$s$^{-1}$.  The solid contours give the position of
    the I--front at 0.15, 0.25, 0.38, and 0.57\,Myr after the quasar
    has switched on. The underlying grey-scale image indicates the
    initial HI density field.}  \label{fig:Ifront} \end{center}
\end{figure}
Consider the simple case of only absorption. Then across a
computational cell where we assume the density of absorbing material
to be constant the outgoing photon number flux is simple $e^{-\tau}$
times the incoming one. Hence the number of absorbed photons must be
$(1-e^{-\tau})$ times the incoming flux. So one can compute the number
of photoionizations per second by adding all these $(1-e^{-\tau})$
terms for the rays that pass this cell. Now by definition we ensure
that the number of photoionizations will always equal the number of
photons absorbed. As a consequence this method propagates ionization
fronts at always the correct speed {\sl independent} of
resolution. This is a highly desirable feature of any method of
multi-dimensional radiative transfer. This control of accuracy comes
at high computational cost. In this method the $1/r^2$ drop in the
photon flux in an optically thin region around the source is captured
by the simple fact that many more rays traverse through cells near the
source than cells further away. Obviously a large amount of
computational time is wasted on computing the flux in such optically
thin cell where it would simply be given by $I(0)/r^2$. This can be
overcome as is discussed below. However, this method nevertheless can
be used for a variety of realistic cases. This can be seen from Figure
1 and 2. Both of them employed the ray-tracing of Abel, Norman and
Madau (1999) and are here shown as illustration of the practicality
of this method.
\subsection{Ionization Front tracking}
Let us quickly side-track to point a simple way of solving a specific
problem. If one is interested in the propagation of a R-type
ionization front in a static medium it suffices to integrate the jump
condition
\begin{equation}\label{equ:jump}
n_{HI}(r) \frac{dR}{dt} = \frac{F(0)}{4\pi R^2} - \int_0^R \alpha(T(r))
n_{HII}(r) n_e(r) dr, 
\end{equation}
where $R$, $\alpha$, and $F(0)$ denote the radius of the I-front, the
recombination rate--coefficient, and the ionizing photon number
luminosity, respectively. Dividing by $n_{HI}$ we can integrate
equation (\ref{equ:jump}) explicitly along rays\footnote{where one
uses the raytracing technique of Abel, Norman, and Madau (1999)} and
find the time at which the ionization fronts arrives at a given
cell. Storing the arrival time in a 3D array allows one to investigate
the time dependence morphology of ionization front by simply taking
iso-surfaces on this array of arrival times. Such data is also
interesting to compute how many ionizing photons are used to ionize a
given volume in a static case, etc.. To get the full time evolution of
the ionization front of one source on a 128$^3$ numerical grid
requires $\sim$ one minute computation (wall clock) on a
workstation. 

\begin{figure}[t]
  \begin{center}
    \leavevmode
    \includegraphics[width=7.cm]{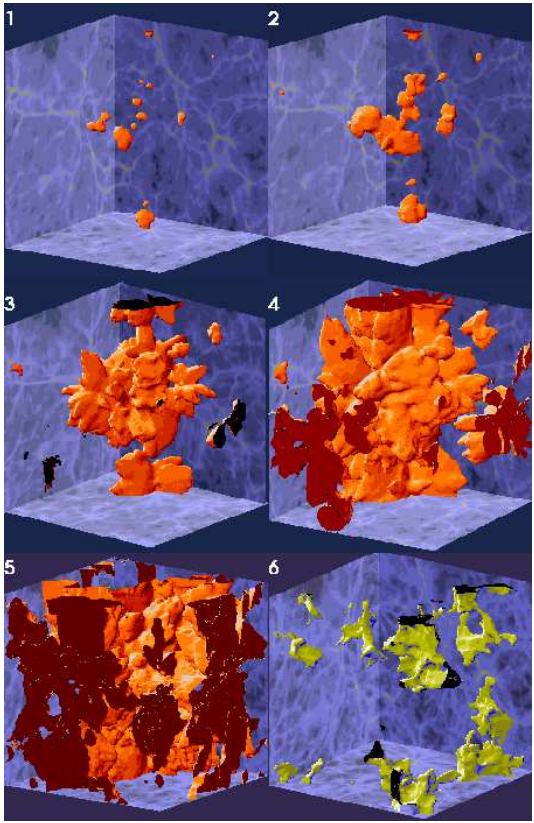}
    \caption{Visualization of a time series of a 3D radiative transfer
     simulation. Here we assumed 30 sources with an ionizing
    luminosity proportional to their halo mass that switch on
    simultaneously at redshift 7. The density distribution is take
    from a standard cold dark matter Lyman alpha forest simulation in
    a periodic box of 2.4 comoving Mpc carried out by Bryan et al. 
    1999. Hence the box corresponds to proper 300 kpc across. Six
    different panels are shown. Each slice at the 
    cube edges illustrates the log of the hydrogen density. The
    iso--surfaces visualize the boundaries of ionized regions. The
    last panel, number six, illustrates remaining neutral islands that
    have not been ionized yet.}
    \label{fig:cartoon}
  \end{center}
\end{figure}

\subsection{Computer Graphics}

In many fields as e.g. biomedical imaging interactive volume rendering
of 3 dimensional data is highly desirable. A lot of effort went into
designing fast algorithms that yield optical depths from a light
source and to the observers eye. Not any such method will be suitable
for application in astrophysics in particular one needs to worry about
exact photon (energy) conservation. However, imagine one has a method
that gives the optical depth to a source at every point in the
computational volume. For the case of pure attenuation one then also
knows the photon number flux (photons per second per area) everywhere
from 
\begin{equation} 
\vec{F}(\vec{r}) = \frac{F_{source}(0)}{|\vec{r}|^2}\  e^{-\tau(\vec{r})}
\frac{\vec{r}}{|\vec{r}|}, 
\end{equation}
where $\vec{r}$ denotes the vector from the source to the point of
interest. 
Now from the obvious 'discontinuity equation':
\begin{equation}
\nabla \cdot \vec{F}(\vec{r}) -  \dot{n_{HI}} = 0 
\end{equation}
one can ensure the number of photoionizations $\dot{n_{HI}}$ to
be computed self--consistently independently of resolution. 
Such a method has been implemented and tested by Abel and Welling
(2000) and found to give speed ups in excess of a factor hundred as
compared to Abel, Norman and Madau (1999) in the limit of large
ionized regions.

\subsection{Moment Methods}

The methods presented above focus and the correct implementation of
radiative transfer for point sources. However, ideally we also want to
be able to treat regions of diffuse emission as it arises e.g. due to
bremsstrahlung and recombination radiation.  In Norman, Paschos and
Abel (1998) we have outlined a possibility approach to treat point
sources and diffuse radiation by means of a variable Eddington tensor
formalism.  Although we have significantly improved on some
ingredients of this method (as e.g. we derived an analytic expression
for the Eddington tensors in the pre--overlap stage) we have not
succeeded as yet in constructing a stable implementation.

\section{Concluding Remarks}

For the applications to numerical cosmology some of the methods of 3D
radiative transfer discussed above will have to be combined. I-front
tracking is useful to initialize the environment of new sources. The
methods drawn from Computer Graphics can be used to compute accurate
boundary conditions for the moment methods that are the most promising
in the limit of many sources. A number of interesting problems still
will need to be solved before cosmological radiation hydrodynamics can
become a standard tool for the study of the formation and evolution of
structure in the universe. However, the existing techniques should be
employed for the study of interstellar problems in which only few
sources are of interest. Planetary Nebulae and HII regions are ideal
candidates for such three-dimensional radiation magneto--hydrodynamic
modeling.

\acknowledgements I am greatful to my collaborators Pascal Paschos,
Mike Norman, Piero Madau, Aaron Sokasian, Lars Hernquist, and Joel
Welling for all the fun we are having in devising these new approaches
and learning the physics. Part of this work was supported by NASA ATP
grants NAG5-4236 and NAG5-3923.

\end{document}